![Plat_Forms — The web development platform comparison]

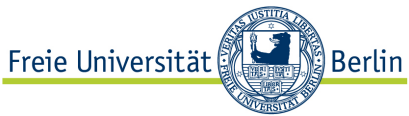 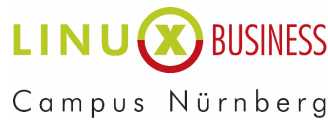 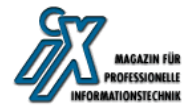

# Plat_Forms — a contest:
# The Web-development platform comparison

Lutz Prechelt (prechelt@inf.fu-berlin.de)

Freie Universität Berlin
Institut für Informatik
2006-10-12, Version 9
(Technical Report B 06-11)

**Abstract**

"Plat_Forms" is a competition in which top-class teams of three programmers compete to implement the same requirements for a web-based system within 30 hours, each team using a different technology platform (Java EE, .NET, PHP, Perl, Python, or Ruby on Rails). The results will provide new insights into the real (rather than purported) pros, cons, and emergent properties of each platform. The evaluation will analyze many aspects of each solution, both external (usability, functionality, reliability, performance, etc.) and internal (structure, understandability, flexibility, etc.).

*Hint for impatient readers:* Sections 1 and 2 (together just one page long) provide all of the overview information. The rest of the document contains the details.

**Table of contents**

Changes:
- Version 8, 2006-10-05: Initial public version
- Version 9, 2006-10-12: Added Perl explicitly to the list of platforms, corrected date mistake ("2006" must be "2007")

-

# 1. Overview

Software development platforms for web applications (such as Java EE, .NET, PHP, Perl, Python, Ruby on Rails, etc.) are among the most critical factors of development productivity today. The pros and cons of the various platforms are by-and-large known in principle, but how the pros trade off against the cons in one platform and how that compares to another platform is the topic of quasi-religious wars only, not a subject of objective analysis, as almost no data is available that allows such direct comparison.

Plat_Forms is a contest that will change this. It will have top-class (and hence comparable) teams of 3 programmers implement the same specification of a web-based application under the same circumstances and thus generate a basis for objective comparison of the various characteristics that the platforms generate.

In just 30 hours, the teams will implement as much of the requested functionality as they can and at the same time optimize the usefulness of the resulting system (functionality, usability, reliability, etc.), the understandability of the code, the modifiability of the system design, the efficiency and scalability.

The contest will be conducted at the conference "Open Source Meets Business" in Nürnberg, Germany (www.open-source-meets-business.com) on January, 25-26, 2007. At the end of 30 hours, the teams hand over their source code and a DVD containing a turnkey-runnable VMware configuration of their system.

These items will then be subject to a thorough evaluation according to scientific standards with respect to the criteria mentioned above. As most of the results cannot be quantified and many cannot even be ranked in a neutral fashion, there will be no one-dimensional result ranking of the systems. Rather, there will be an extensive report describing the findings. Depending on the results, the organizers may or may not declare one or a few of the systems and teams winners with respect to one particular criterion (and do so for some or all of the criteria).

The contest is organized by Freie Universität Berlin (www.fu-berlin.de), Linux Business Campus Nürnberg (www.lbcn.de), and 'iX magazine (www.ix.de).

# 2. How the contest will proceed

- before 2006-11-30: Teams apply for participation in the contest as described under "How to apply" below
- 2006-12-14: Teams are notified whether they will be admitted to the contest. At most three teams per platform will be admitted, a total of 15 teams.
- 2007-01-24: Teams set up their development environments at the contest site. For details, see "Infrastructure" below.
- 2007-01-25, 9:00: The contest starts. The organizers will explain (in a presentation format) the requirements of the system to be developed, will hand out a short document containing a few more details, and will answer any immediate questions that may arise. For details, see "The task" below.
- 2007-01-25, 10:00: The teams start developing the software using their favorite platform and tools. Reusing existing software and reading the Web is allowed, getting external help is not. For details, see "Rules of behavior" below. The teams are asked to make intermediate versions accessible for user feedback, see "Semi-public preview" below
- 2007-01-26, 16:00: The teams stop developing software and hand over their result in the form of a VMware image of a server machine. For details, see "Results hand-over" below. Teams that believe they have reached the best cost-benefit ratio of their development before the allotted time is over are allowed to hand-over their results earlier and will have a shorter work time recorded.
- 2007-02-02, 16:00 (that is, 15:00 UTC): The teams submit post-hoc design documentation. For details, see "Explaining what you did" below.

- 2007-02-02: Evaluation of the systems starts. It will investigate all categories of quality criteria, both internal and external. For details, see ["Evaluation and winning"](#)
- 2007-05: Results of the contest will be presented. The details of when, where, and how are still to be determined. See ["What's in for the teams and their organizations?"](#) for what is known already.

## 3.  Do we really need another contest?

Absolutely. But it is not "another", it is the first of its kind.

### 3.1.  The platform decision situation

Every year, several hundred million dollars are spent for building the type of application mentioned above, yet nobody can be quite sure in which cases which platform or technology is the best choice. Quasi-religious wars prevail.

Some platforms are often claimed to yield better performance than others, but nobody can be quite sure how big the difference actually is.

Some platforms are often claimed to yield higher productivity in initial development than others, but nobody can be quite sure how big the difference actually is.

Some platforms are often claimed to yield better modifiability during maintenance than others, but nobody can be quite sure how big the difference actually is -- or if it really exists at all.

So as a program manager one can almost consider oneself lucky if a company standard prescribes one platform (or if expertise is available only for one) so that the difficult choice needs not be made.

However, that means that many (if not most) projects may use a sub-optimal platform -- which sounds hardly acceptable for an industry that claims to be based on hard knowledge and provable facts.

### 3.2.  What we need

What we need is a direct comparison of the platforms under realistic constraints: a task that is not trivial, constrained development time, and the need to balance all of the various quality attributes in a sensible way.

### 3.3.  Why it is still missing

So if such a comparison is so important, why has nobody done it yet?

Because it is difficult. To do it, you need:
- participant teams rather than individuals, or else the setting will not be realistic;
- top-class participants, or else you will compare them rather than the platforms;
- a development task that is reasonably typical, or else the result may not generalize;
- a development task that is not too typical or else you will merely measure who of the participants happened to have a well-fitting previous implementation at hand;
- participant teams that take the challenge of implementing something on the spot that they do not know in advance;
- the infrastructure and staff to accommodate and supervise a significant number of such teams at once;
- an evaluation team that is capable of handling a heterogeneous set of technologies;
- an evaluation team that dares comparing these fairly different technologies (in some respects almost apples and oranges) in a sensible yet neutral way;

## 3.4. What we've gotten so far

For these reasons, all previous platform comparisons were very restricted. Several, such as the c't Database Contest or the SPEC WEB2005, concentrate on one quality dimension only (typically performance) and also provide participants with unlimited time for preparing their submission. Others, such as the language comparison study by Prechelt, are broader in what they look at and may even consider development time, but use tasks too small to be relevant.

Plat_Forms will be a big step forward for everybody's future platform selection information base.

## 4. How to apply

At most three teams per platform will be admitted to the contest. It is *not* the purpose of the contest to compare the competence of the teams; we will therefore strive to get the best possible teams for each platform to make it more likely that significant differences observed in the final systems can be attributed to the technology rather than the people.

Teams interested in participating please apply by sending a *Request for Admittance* in the form of a 3-page document as described on http://www.plat-forms.org. Only one application per platform is allowed for each home organization (company). Teams must agree that the result of their work (but not frameworks etc. that they bring along) released under the GNU General Public Licence V.2 (GPL) or the modified BSD license.

From among the applications, teams will be selected so as to maximize their expected performance. The selection process is performed with the help of a contest committee for which we will invite representatives from an owner or protagonist organization of each platform (e.g. Sun, Microsoft, Zend, Zope).

## 5. Infrastructure

The following information is preliminary. We will provide an update with possible modifications of some details two weeks before the contest.

At the contest, you will be provided with roughly the following infrastructure:

- Electrical energy (235 V, 5 A max., separately fused, german-style Schuko socket, see http://en.wikipedia.org/wiki/Schuko)
- 2 tables 200x80 cm
- 3 chairs
- One flipchart stand with plenty of flipchart paper
- 4 meters length of wall
- 4 meters by 4 meters of floor space
- an internet connection, via an RJ45 connector serving 100 MBit Ethernet. The available bandwidth is not yet known, bandwidth management is probably on a best-effort basis.
- sufficient food and drink

All teams will work in one single large room.

Things you need to bring yourself:

- computers, monitors, keyboards, mice for 3 developers,
- a server computer
  - must be operable in stand-alone mode, without network connection,
  - must host at least the turnkey configuration of your final system under VMware
- network cables, network hub,
- printer, printer paper,
- pens, markers, scissors, adhesive tape,

- perhaps desk lamp, pillow, inflatable armchair etc.
- coffee mug,
- backup coffee mug.

# 6. The task

We will obviously not tell you right now what the development task will be. However, here are some considerations that guide our choice of task:

- It will be a web-based application with both a browser-based front end and a SOAP web service interface. The browser interface must be compatible with all major browsers (IE, Firefox, Opera, Safari).
- It will require persistent storage of data. It will not require integration with external systems or data sources.
- It will neither be a web shop (or other highly standard type of application, say, static-content management) nor something entirely exotic and unprecedented. The task will be chosen such as to make reuse of large portions of existing systems unlikely, but reuse of smaller pieces possible.
- In order to allow for a broad assessment of a platform's characteristics, the task will not be a create-read-update-delete database application only, but will involve other aspects as well. Such things *might* be for example algorithmic processing, data-driven graphics, audio handling, etc. The complexity of these requirements will be modest, so that they can be solved without specialist knowledge.

In your solution you should strive for a good balance of all quality attributes. This includes usability and robustness.

# 7. Rules of behavior

## 7.1. What is allowed

During the contest you may:

- Use any language, tool, middleware, library, framework, and other software you find helpful (just please mention as many of these as you can foresee in your application).
- Reuse any piece of any pre-existing application or any other helpful information you have yourself or can find on the web yourself. Anything that already existed the day before the contest started is acceptable.
- Use any development process you deem useful.
- Ask the organizer (who is acting like a customer) any question you like regarding the requirements and priorities.

## 7.2. What is not allowed

During the contest you may not:

- Disturb other teams in their work.
- Send contest-related email to people not on your team or transfer the requirements description (or parts thereof) to people not on your team.
- Have people from outside of your team help you or "reuse" work products from other teams. There are two exceptions to this rule: you may use answers of the customer and user-level preview feedback as described below.

# 8. Semi-public review

During the contest, teams will be able to obtain feedback from the conference participants if they wish to do so. For this purpose, the team should open their test system on their VMware team server for public access.

The organizers will put up a Blog where the teams can announce their release plan (if any), releases, and access URLs, and where the conference participants can comment on the prototype systems regarding functionality, defects, usability etc. The teams are allowed to use this user-level feedback for improving their system. They are not allowed to take or use information from conference participants that is on the code level.

Depending on whether we manage to provide suitable firewalling and bandwidth management, "the public" may either mean the local network at the conference site (open to the conference participants only) or the whole Internet. This is so far undetermined.

## 9. Results hand-over

The technology used for building the systems in the contest will be very heterogeneous. It would therefore be impractical for the contest organizers to try to execute them from source code alone, not to speak of obtaining similar behavior in a performance test.

We thus require each team to deploy their solution on a virtual server that is running under VMware Server 1.0 (see www.vmware.com). The image file of this virtual server will be handed over at the end of the contest by means of a single DVD-R created by the respective team themselves. This means the image must be smaller than 4.7 GB, which should easily be possible, because the virtual server does not need to have any application software installed beyond your contest solution and the infrastructure software that it uses.

Beyond the image file, the DVD needs to contain a second file that is an archive (zip or tar.gz) containing a snapshot of all source artifacts (source code, build files, database initialization scripts, configuration files, CSS files, etc.) that are part of the solution. The contents of this archive must be sufficient in principle to recreate your solution from scratch, given the infrastructure software (such as operating system, build tools, DBMS, application server etc.) Furthermore, a third file must contain your source code version archive so the organizers can analyze some aspects of the development process.

At the time of the DVD handover, the teams will also send a cryptographic fingerprint of the image file and of the archive files to the organizers by email, so that a replacement DVD can be accepted should the original DVD-R fail to be readable.

The virtual server must be able to run in stand-alone mode, that is, in a network where no other servers and services (other than DHCP) exist. We will ask you to provide us with necessary information about the virtual server, such as usernames/passwords, etc.

During the evaluation, we will run each team's virtual server on a dual-CPU server with a two-disk SCSI RAID-0. The virtual server will be allotted 3 GB of RAM.

## 10. Explaining what you did

Both source code and build/configuration/deployment of your system are fixed at server hand-over time. However you will be able to prepare and submit a document afterwards that shortly explains the following points:
- the architecture of your system
- your approach to development (priorities, implementation orders etc.)
- the rationale of each important design decision you have identified
- etc.

This document will be an important contribution towards a fair and thorough evaluation of your system, because without it the evaluation team will have a hard time judging many of the things it will get to see.

We will provide you with a template for this document.

## 11. Evaluation and "winning"

We will attempt to evaluate all of the following aspects of the system and its development:
- External product characteristics: functionality, ease-of-use, resource usage, scalability, reliability, availability, security, robustness/error checking, etc.
- Internal product characteristics: structure, modularity, understandability, modifiability (against a number of fixed, pre-determined scenarios), etc.
- Development process characteristics: Progress over time, order and nature of priority decisions, techniques used, etc.

The details of this evaluation will be determined once we get to see the systems that you built. The evaluation will be performed by the research group of Professor Lutz Prechelt, Freie Universität Berlin.

We will not compare all systems in one single ranking by some silly universal grading scheme. Rather, we will describe and compare the systems according to each aspect individually and also analyze how the aspects appear to influence each other.

Therefore, there may be "winners" of the contest with respect to individual aspects (or small groups of related aspects) where we find salient differences between the platforms or the teams. However, there will not be a single overall winner of the contest.

## 12. What's in for the teams and their home organizations?

So why should you participate in the contest if you cannot win it?

Two reasons:

1. Category "riches and beauty": We *will* award monetary prices, just not across platforms. However, we will nominate a best solution among the three solutions on each individual platform.

2. Category "eternal fame": The detailed evaluation will provide the organizations of the well-performing teams and platforms with some of the most impressive marketing material one can think of: concrete, detailed, neutral, and believable.

## 13. Further information

For details regarding
- the organizers of the contest,
- the sponsors,
- the Contest Committee roles and members,
- the Request for Admittance process and document,
- and other information

please visit http://www.plat-forms.org